# Considerations for missing data, outliers and transformations in permutation testing for ANOVA, ASCA(+) and related factorizations

Oliver Polushkina Merchanskaya[1*], Michael D. Sorochan Armstrong[1*], Carolina Gómez Llorente[2], Patricia Ferrer[3], Sergi Fernandez-Gonzalez[3], Miriam Perez-Cruz[3], María Dolores Gómez-Roig[3], José Camacho[1]

[1] Research Centre for Information and Communication Technologies (CITIC-UGR), University of Granada, Spain

[2] Institute of Biosanitary research ibs.GRANADA, 18012 Granada, Spain. Department of Biochemistry and Molecular Biology II, Faculty of Pharmacy, Campus Universitario de Cartuja, 18071, Granada, Spain. Institute of Nutrition and Food Technology "José Mataix", Biomedical Research Centre, University of Granada, 18100, Granada, Spain. Biomedical Research Centre in Physiopathology of Obesity and Nutrition (CIBERObn), CB12/03/30038 Institute of Health Carlos III (ISCIII), 28029, Madrid, Spain

[3] BCNatal, Barcelona Centre for Maternal-Fetal and Neonatal Medicine, Hospital Sant Joan de Déu and Hospital Clínic, University of Barcelona, 08950, Barcelona, Spain; Institute of Research Sant Joan de Déu, 08950, Barcelona, Spain; Primary Care Interventions to Prevent Maternal and Child Chronic Diseases of Perinatal and Developmental Origin (RICORS), RD21/0012/0003, Institute of Health Carlos III (ISCIII), 28029, Madrid, Spain

**Abstract**

Multifactorial experimental designs allow us to assess the contribution of several factors, and potentially their interactions, to one or several responses of interests. Following the principles of the partition of the variance advocated by Sir R.A. Fisher, the experimental responses are factored into the quantitative contribution of main factors and interactions. A popular approach to perform this factorization in both ANOVA and ASCA(+) is through General Linear Models. Subsequently, different inferential approaches can be used to identify whether the contributions are statistically significant or not. Unfortunately, the performance of inferential approaches in terms of Type I and Type II errors can be heavily affected by missing data, outliers and/or the departure from normality of the distribution of the responses, which are commonplace problems in modern analytical experiments. In this paper, we study these problem and suggest good practices of application.

[*]Both authors contributed equally

# 1 Introduction

Over the last few decades there has been a significant increase in the quality and quantity of biological and chemical data, owing to advances in analytical instrumentation. In particular, the application of omics technologies from various biological domains such as genomics, transcriptomics, proteomics, or metabolomics, has proven to be a powerful approach in the clinical field [1]. The relative dimensionality of biochemical features to samples is typically quite high for most omics data. When dealing with these multivariate responses in complex datasets, it is often the case that there are some variables that are highly correlated with the interesting phenotypes being studied, and others that are not [2]. A way of simplifying the interpretation of the results is to apply a series of univariate ANalysis Of Variance (ANOVA) tests [3]. However, univariate analyses only examine a single response and do not consider correlations between different responses as they relate to the hypotheses being tested. Thus, possible multivariate relationships that best describe the differences in the observable properties of the samples may be missed. In these cases, multivariate extensions of ANOVA are generally preferred.

A multivariate version of ANOVA that is gaining momentum is ANOVA Simultaneous Component Analysis (ASCA) [2]. ASCA combines the variance factorization and inference capabilities of ANOVA with the exploratory power of Principal Component Analysis (PCA). In a three-step process, ASCA performs a matrix factorization based on the factors and interactions of the experimental design, performs statistical inference to compute p-values associated to the entire response matrix, and creates visualizations based on PCA for each of the significant factors, interactions, or their combined effect. These factor/interaction subsets can be tested for evidence of significance by comparing a statistic associated with the different components of the experimental design, against a null distribution, often derived from permutation testing [4]. In both univariate ANOVA and the state-of-the-art version of ASCA [5] (often referred to ASCA(+)), the matrix factorization in the first step follows an identical approach based on the General Linear Model (GLM).

A ubiquitous problem while analyzing data, especially in clinical trials, is the handling of missing data. Missing data has compromised the inference and conclusions in clinical trials [6]. There is no universal method for handling missing data, since the number of missing values, and external considerations such as the relative distribution of missing values across an experimental design, must be taken into account.

Another problem for the application of an inferential test is that its assumptions must be confirmed, including the independence of experiments and the (quasi-)normality and homoscedasticity of residuals. Independence should be guaranteed from the experimental design. If the other assumptions are not met, a variance-stabilizing transformation [7] or a less restrictive test that departs from these assumptions altogether is necessary. Box-Cox transformations represent a series of operations that map data to a distribution closely approximating a normal distribution. In cases where the data presents several extreme values, a Box-Cox transform will reduce their influence on the measured variance of the data, which makes it easier to compare across multiple variables. Distribution-free testing improves the performance of significance tests on models whose residuals are non-normally distributed but contain otherwise exchangeable data.

In this paper, we study the problem of handling missing data and non-normal distributions, including the presence of outliers, in multiple responses of an experimental design in the context of (multiple) ANOVA and ASCA, and suggest good practices of application. To motivate our choice of good practices, we will use a mixture of simulated experiments and real data. Different simulation approaches are described in the remainder of the paper where needed. For the reproducibility of the results, the code of these simulations can be found in the repository at github.com/CoDaSLab/glm_factorization_2024.

The rest of the paper is organized as follows. Section 2 introduces the ANOVA/ASCA(+) factorization using the GLM approach. Section 3 discuses inferential procedures, including parametric ANOVA, permutation testing and multiple-testing correction approaches. Section 4 presents the missing data imputation problem and suggested solutions in the context of the paper. Section 5 does the same for outliers management and variance-stabilizing transformations. Section 6 applies suggested solutions to a real case study. Section 7 provides a general discussion and Section 8 brings the conclusions of the work.

## 2   ANOVA/ASCA(+) factorization through General Linear Models

In the remainder of the paper, we identify scalar elements with lowercase letters, array dimensions with uppercase letters, vectors with bold lowercase letters, and matrices with bold uppercase letters.

Consider an $N \times 1$ vector **x** of experimental data, a.k.a. a (response) variable, which represents the response of a experimental design with two factors, A and B, an which can be factorized according to:

$$\mathbf{x} = \mu + \mathbf{x}_A + \mathbf{x}_B + \mathbf{x}_{AB} + \mathbf{e} \tag{1}$$

where $\mu$ is the vector mean, $\mathbf{x}_A$, $\mathbf{x}_B$, $\mathbf{x}_{AB}$ represent the factorizations of the response according to the experimental factors and interaction, and $\mathbf{e}$ is the residual vector describing variance unaccounted for by the model. Each experimental factor or interaction is encoded according to a coding matrix, $\mathbf{C}$, using sum coding [5] or an alternative coding scheme [8]. The factorization for an arbitrary combination of factors and/or interactions is calculated in GLM from the least squares regression of the response on the encoded experimental factors:

$$\mathbf{\Theta} = (\mathbf{C}^T\mathbf{C})^{-1}\mathbf{C}^T\mathbf{x} \tag{2}$$

with $\mathbf{\Theta} = [\mu, \mathbf{\Theta}_A, \mathbf{\Theta}_B, \mathbf{\Theta}_{AB}]$ which minimize the sum-of-squares of the residuals:

$$\mathbf{e} = \mathbf{x} - \mathbf{C}\mathbf{\Theta} \tag{3}$$

We can easily extend Eqs. (1)-(3) to the case of multivariate responses encoded as a series of vectors concatenated horizontally in matrix $\mathbf{X}$, which can be similarly factorized:

$$\mathbf{X} = \mathbf{C}\mathbf{\Theta} + \mathbf{E} = \mu + \mathbf{X}_A + \mathbf{X}_B + \mathbf{X}_{AB} + \mathbf{E} \tag{4}$$

where $\mu$ is a column-wise vector of means, and the matrices $\mathbf{X}_A, \mathbf{X}_B, \mathbf{X}_{AB}$ represent the factorization with respect to their subscripted designator, and $\mathbf{E}$ is the matrix of residuals. Eq. (4) represents the first step in ASCA+ [5], and it is equivalent to performing Eq. (1) for each response in $\mathbf{X}$ (each column vector) independently.

## 3 Inferential Procedures

Inferential procedures associated with ANOVA/ASCA(+) allow us to assess whether the contribution of the individual factors and their interactions to the response(s) demonstrate evidence of statistical significance. In standard univariate ANOVA it is possible to use parametric methods based on typical assumptions (such as normality) [9]. However, in the context of more than just a few responses, or to assess any potential violation of assumptions, numerical methods [10] are a viable alternative. Arguably the most popular approach for statistical inference in ASCA is permutation testing [11]: a resampling method that transforms ASCA into a distribution-free approach that is more flexible than parametric ANOVA. Permutation testing can also be used as the inferential method in ANOVA.

In this paper, our focus is on multiple responses, for which inference results may be approached either to each response independently, making multiple-testing correction, or on a multivariate basis. For the sake of comparability between parametric and permutation tests, we opt for univariate tests, but we expect our conclusions to be generally applicable.

3.1  Parametric Methods

Parametric ANOVA is based on the ratio of the mean sum-of-squares (MS) between and within the levels/cells of a factor/interaction. This ratio is contrasted with the F distribution with the suitable degrees of freedom (DoFs). The MS is the sum-of-squares (SS) of a factor or interaction normalized by its corresponding DoFs. The DoFs of a factor is the number of levels minus one. The DoFs of an interaction is the product of the DoFs of its factors. The DoFs of the residuals is the total DoFs minus the DoFs of all factors and interactions in the model. The total DoFs is the number of observations minus 1. To test the significance of each factor and interaction, the correct F ratio is determined, taking into account the nature (fixed or random) and relationships (relative orders of crossed/nested factors and of interactions) among the elements of the experimental design [9], [10].

Parametric tests assume that the observations are adequately described by the model (e.g., Eq. (1)), and that the residuals are normally and independently distributed with zero mean and constant variance (homoscedasticity) [9]. It is always possible to check these assumptions in the data: i) residuals can be tested for normality and/or visualized with normality plots, ii) residuals can be visualized in terms of the levels of the factors to find indications of heteroscedasticity; and iii) residuals can be plotted in time to find dependency traces. Violation of independence assumptions is a potentially serious problem regardless of the inference approach, and one which should be avoided from the experimental design. Moderate departures from normality and from homoscedasticity are expected to affect more ANOVA models with random effects than with only fixed effects. Heteroscedasticity is also more relevant for unbalanced designs. Quite often, these departures are handled by transformations, with the goal of bringing the distribution of the residuals closer to normality and stabilizing the variance.

3.2    Permutation testing

Permutation testing is a distribution-free testing approach, especially useful in instances where the normality assumption is not fully satisfied. Permutation tests permute the rows (experimental units) of the data a number of times and measure the frequency in which the statistic measured on the permuted data (often the SS of a factor/interaction or a suitable F-ratio) exceeds the measured statistic on the un-modified data. There are several variants of permutation testing, from the exact test, permutations of the raw data, permutations on the residuals of the reduced model, and permutations of the full model residuals. Permutation testing of the raw data is most commonly used in the context of ASCA [11], but may often be suboptimal in univariate tests [10]. This observation conflicts with similar experiments for multivariate responses [12].

3.3    Univariate tests with multiple-testing correction

As the number of tests increases, the likelihood of a test falsely rejecting the null hypothesis (Type I error) beyond a certain significance level increases. Thus, when the experiment presents multiple responses, univariate inferential results need to be corrected by multiple-testing corrections, like the Family-Wise Error Rate (FWER) and the False Discovery Rate (FDR). The Bonferroni correction is one FWER technique that simply divides the significance level by the total number of tests. This lowers the Type I error but is a rather conservative correction that can increase the likelihood of false negatives, or Type II errors. A more powerful multiple-testing correction is the FDR [13]. Storey-Tibshirani $q$-values are a slight modification of the FDR that represents the proportion of $p$ values that are expected to correctly reject the null hypothesis, accounting for the proportion of tests already anticipated to do so [14], [15].

## 4    Missing Data imputation in GLM

One simple way to handle missing data is by using mean replacement imputation techniques. Among these techniques, the simplest approach is unconditional mean replacement (UMR), in which a missing value for a certain response is imputed as the mean of all available measurements for that response. An alternative approach is conditional mean replacement (CMR). In GLM, CMR imputes a missing value for a certain response (e.g., $x$ in Eq. (1)) in an experimental unit as the mean of that response computed only from the replicates in the same cell, that is, other experimental units with the same levels.

4.1  Impact of missing data imputation in the GLM factorization

Mean replacement imputation techniques are generally regarded as simplistic in the multivariate domain, since they do not consider the relationships among (response) variables [16]. Yet, this may be seen as an advantage in the context of GLM, since no predictive (and to some extent artificial) variance is transferred among the set of responses, potentially impacting the statistical inference results in an uncontrolled manner, difficult to validate. Table 1 illustrates the average behavior of UMR and CMR in 10 simulated examples. Each example simulates two factors (Factor A: 4 levels, Factor B: 3 levels), one of them significant, and with 400 responses, where 5% of missing data is induced. The considered factorization model is:

$$\mathbf{X} = \mathbf{\mu} + \mathbf{X}_A + \mathbf{X}_B + \mathbf{E}_{A,B} \tag{5}$$

We also included the imputation by Trimmed Score Regression (TSR) [17], a popular but simple multivariate imputation method. The SS of the original, complete data is 21038. After inducing missing data, the SS of available data drops by approx. 1000 (approx. 5% of the total). Note this will happen in any real example: the SS of the available data is by definition lower or equal than that of the complete (original) data, just because less data items are considered since some are missing. If we look at the outcomes of the imputation methods, UMR and TSR produce the closest SS factorizations to that of the available data. CMR achieves the best approximation to the original data, but this result is subject to the specific missing data generation strategy and the fact that Factor A is statistically significant. However, we do not expect that this is a general result, as shown in Section 4.2.

**Table 1**
Average Sum of Squares for 10 simulations with two factors, one significant, and no significant interaction. Missing data (5%) is artificially generated in the data and imputed with Unconditional Mean Replacement (UMR), Conditional Mean Replacement (CMR) and Trimmed Score Regression (TSR).

|  | **Original** | **Available** | **UMR** | **CMR** | **TSR** |
|---|---|---|---|---|---|
| **Factor A** | 14749 | 13964 | 13260 | 14756 | 13723 |
| **Factor B** | 63 | 98 | 92 | 68 | 91 |
| **Residuals** | 6226 | 5883 | 6852 | 6154 | 6421 |
| **Total** | 21038 | 19946 | 20204 | 20979 | 20234 |

In this paper, we opt for mean replacement techniques so that we maintain the factorization of the data intrinsically univariate. Note that the GLM procedure introduced in Section 2 is multi-way (so that several factors and levels are jointly considered) but univariate (in terms of the responses), so that we obtain the same factorization for the complete matrix of responses (Eq. (4)) than if we apply it to each individual response (Eq. (1)). With mean replacement techniques, given that we only use column-wise means to impute missing data, GLM remains univariate.

4.2    Impact of missing data imputation in the statistical inference

We expect UMR and CMR to bias statistical inference in favor of non-significance (overestimating p-values) and significance (underestimating them), respectively. This is because UMR replaces each missing element by a general mean, making the differences among factor levels relatively smaller (e.g., compare the SS for Factor A of the missing data set in Table 2, 13964, with that imputed with UMR, 13260), while CMR replaces each missing element by the corresponding average of the experimental cell, making differences bigger (e.g., the SS of Factor A imputed with CMR is 14756).

To provide a better estimate of statistical inference results, we can resort to the permutation testing mechanism combined with one imputation approach, e.g., CMR. We call this approach Permutational Conditional Mean Replacement (or pCMR for short), but the same strategy can be seamlessly extended to other (likely more computationally demanding) imputation approaches. Basically, the idea is to consider the missing data imputation within the permutation testing loop. For each iteration of the permutation test, the missing data are imputed according to the apparent *cell mean* - that is, the mean according to the equivalent rows of *F* with identical encodings. This allows us to incorporate the uncertainty generated due to missing data imputation within the null distribution, and the resulting test statistic accounts for the variance related to the missing data itself.

To compare the effect of the aforementioned missing data imputation methods, we adopt the same simulation approach as before, with two factors, one of them significant. We consider two factorization models, Eq. (5) with two factors, and Eq. (6) with two factors and their interaction:

$$\mathbf{X} = \mathbf{\mu} + \mathbf{X}_A + \mathbf{X}_B + \mathbf{X}_{AB} + \mathbf{E}_{A,B,AB} \tag{6}$$

Note factor B in Eqs (5) and (6) and the interaction in Eq. (6) are non-significant. Comparing the *p*-values on the original data against the *p*-values for data with increasing levels of missing value pollution, it is possible to observe the tendency for each missing value imputation methods to overestimate or under-estimate inference results. The error is defined as:

$$Err(m) = \sum_{i=1}^{n} (p_{obs,i}^m - p_{exp,i})/p_{exp,i} \tag{7}$$

where the observed probability of rejecting the null hypothesis, $p_{obs,i}^m$, at each percentage of missing data ($m$) is measured and normalized relative to the expected probability of rejecting the null hypothesis $p_{exp,i}$ (computed from the complete data) summed over each of the *n* replicates. Eq. (7) is used to preserve the signage of the error term: positive values indicate that the missing value imputation method is overall poorly predictive (high Type II error), and negative errors reveal a tendency for the missing value imputation methods towards Type I errors.

The results in Fig. 1 indicate that for missing values up to 20%, the sum of the normalized differences in the proposed method is well within one standard deviation from the results offered from the original data. For synthetic data with an interaction, the UMR method appears to suffer from more Type I error as the number of missing values increases. And in either case, CMR is shown to be an overly optimistic method for fitting the missing values, with the highest rate of increase in Type I error.

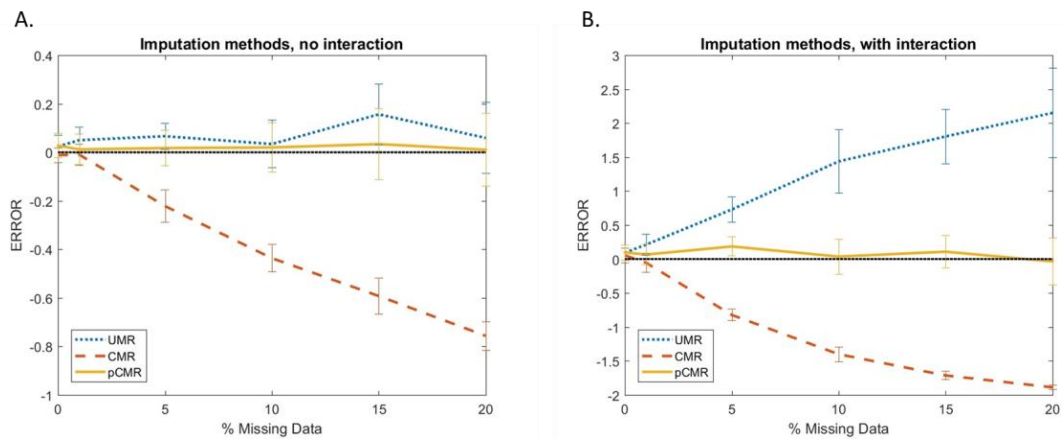

**Fig 1.** Comparison of the relative performance of UMR, CMR, and pCMR as a function of the percent of missing data in a synthetic dataset.

## 5 Transformations and Outlier Detection

As already discussed, a practical problem in ANOVA inference is to handle distributions that deviate from normality. The deviation from normality renders parametric tests prone to Type I and Type II errors. Permutation tests are more robust than fully parametric tests, but they are not immune to severe deviations from normality [10]. Potential solutions are non-parametric tests or performing data transformations towards normality. These solutions have caveats: Non-parametric testing is limited in flexibility in complex designs, and for parametric tests it may be difficult to find the right transformation. The identification of outliers is related to the problem of non-normality. We study those problems in the remainder of this section.

The Box-Cox power transformation family is used on non-negative responses [18]. Another useful transformation is the rank transformation, in which we transform each data value by its corresponding rank within the response. The application of parametric approaches over rank-transformed data nicely approximates the non-parametric techniques [19], while maintaining the flexibility of parametric methods.

In order to evaluate the performance of the Box-Cox and rank transformation in the data set, simulation power curves [12] have been generated for the experimental design in Eq. (6), where residuals in $E_{A,B,AB}$ follow a normal distribution (Fig. 2A), a uniform distribution (Fig. 2B), an exponential distribution then cubed (Fig. 2C), and a normal distribution with one outlier (Fig. 2D). The uniform and exponential cubed distributions have been used elsewhere to emulate the situations of moderate and severe departures from normality, respectively [10]. The simulated experiment has the same structure in factors and interaction as the real data in Section 6. Results provided in the figure are only for factor A (disease/control). We compare the analysis of the raw data (**X**), with the analysis over Box-Cox-transformed data, rank-transformed data, and the combination of the raw data and the rank-transformed data appended in the columns. Power curves should approach the significance level ($\alpha = 0.05$) for null effect size, at the left-hand side, and go up as fast as possible when the effect size is increased. It is important to highlight that these simulated power curves are useful to make relative comparisons among several inference tests over the same data, but the performance of tests on different data should not be compared. Therefore, each single plot in Fig. 2 should be independently interpreted, since the comparison across plots is meaningless (except for Fig. 2A and Fig. 2D, where the only difference is the presence of an outlier).

The main conclusions of the simulation are as follows. Permutation tests on raw data can cope with moderate deviations from normality (Fig. 2B) but not with severe deviations (Fig. 2C) and/or severe outliers (Fig. 2D). The Rank transformation is much more efficient than the Box-Cox transformation in all considered cases, included when the data is indeed normal. This is the only method that is robust to the outlier (compare Fig. 2A and Fig. 2D). Considering these results, our suggestions for good practices are as follows. In real practice, multiway inference can be applied twice, over raw data and over the rank-transformed data. If both results are consistent, it is safe to continue the analysis with the raw data, easier to interpret in, e.g., post-hoc analyses. Otherwise, it may be wiser to continue the analysis with rank-transformed data. In multivariate analyses, this choice may be done per independent response.

Even if the rank transformation is robust to outliers, as expected, it may be a preferable solution to check residuals for the identification of outliers, as it is often done in ANOVA pipelines. That way, we have a better chance that raw and rank transformed data are consistent in the inference results. Furthermore, the identification and interpretation of outliers often provides useful information. In multivariate data, outlier identification is often based on a set of techniques generally referred to as Multivariate Statistical Process Control (MSPC) [20]. A popular approach in MSPC is to combine PCA with a pair of statistics: the D-statistic or Hotelling's T2 statistic for the PCA subspace and the Q-statistic or SPE, which compresses the residuals.

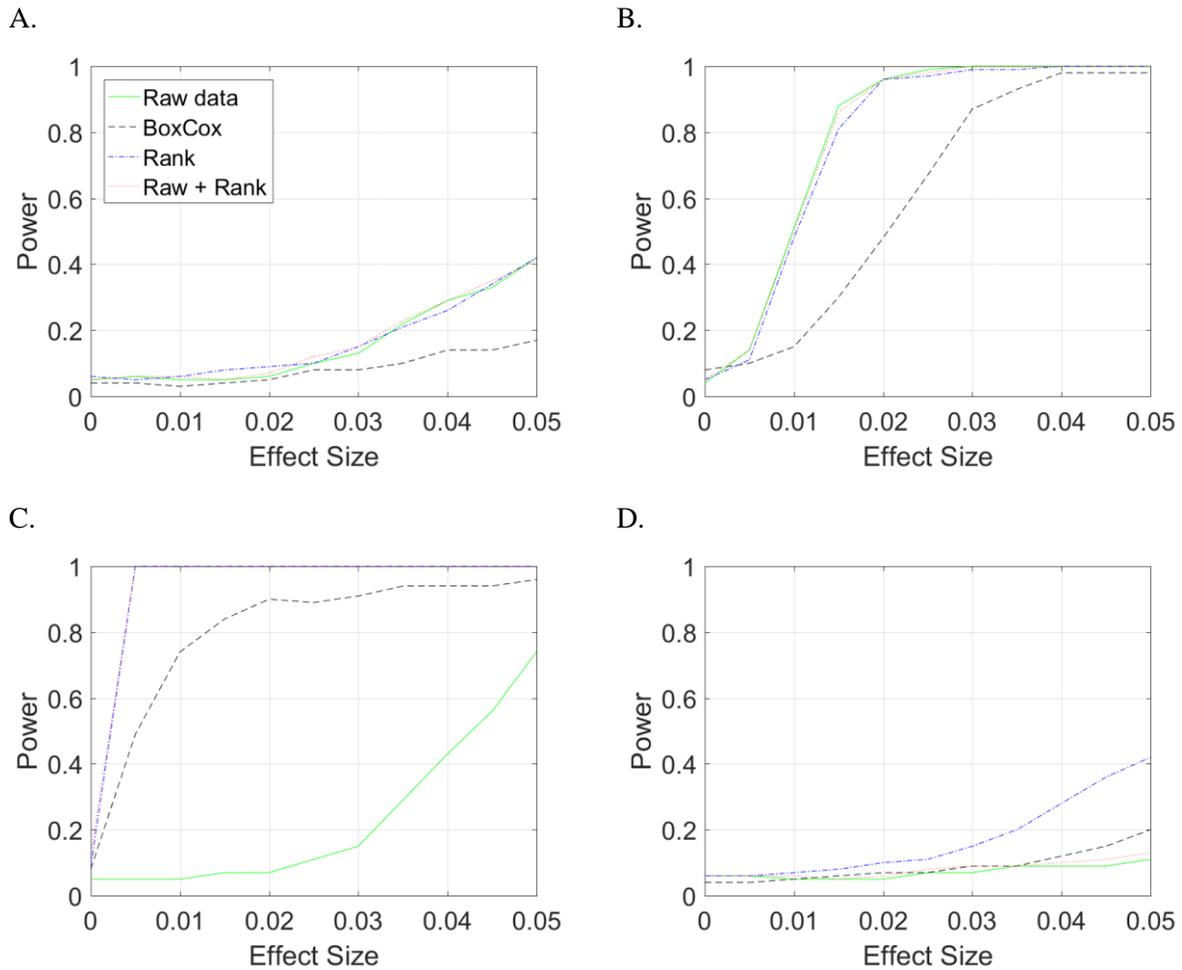

**Fig 2.** Power curves for permutation testing in ASCA for simulated data in which raw, box-cox transformed, rank transformed and raw + rank transformed data are used for factor A (disease/control) where the ASCA residues, E, follow different distributions, A) normal, B) uniform, C) exponential cubed and D) normal with one outlier.

## 6  Results on real data

Data comes from an observational longitudinal study: the BIOCIR study [21]. The objective of the study is to search for inflammatory, metabolomics and metagenomics biomarkers in the late fetal growth restriction (late-FGR) pathology during pregnancy. For the sake of clarity, we only describe results for inflammatory biomarkers, a total of 14 responses. This longitudinal study includes blood samples of two groups of subjects (control and late-FGR pregnancies) along 3 timestamps: pre-labour, after labour time (being those data from the mother) and the newborn's umbilical cord[1]. Analyzed inflammatory biomarkers were determined in maternal plasma samples (pre-labour and labour) and in plasma derived by umbilical cord blood of fetal samples. The studied biomarkers and their acronyms are listed in Table 2.

**Table 2**
Name and acronym for each of the responses in the BIOCIR study.

| **Inflammation biomarkers** | **Acronym** |
|---|---|
| Retinoic Acid Receptor Responder Protein | RARRESS |
| Lipopolysaccharide binding protein | LBP |
| Interferon gamma | IFN-G |
| Interleukin 10 | IL-10 |
| Interleukin 23 | IL-23 |
| Interleukin 6 | IL-6 |
| Interleukin 8 | IL-8 |
| Adiponectin | Adip |
| Resistin | Res |
| Interleukin 15 | IL-15 |
| Vascular endothelial growth factor | VEGF |
| Leptin | Lep |
| Adiponectin/Leptin Ratio | ALR |
| Adiponectin/Resistin Ratio | ARR |

The LBP was determined by simple ELISA whereas the rest of the biomarkers were determined using MILLIPLEXMAP test kits and the Luminex xMAP detection technology [21].

**6.1 Handling missing data**

As the first step, we considered one-way tests for case-control comparisons, one test for each individual response with FDR correction. We did this without checking for outliers, which we will discuss later in the next subsection. We compared the outcome of traditional tests (t-test when normality assumptions were met, Wicoxon tests otherwise) with that of permutation tests. Missing data imputation was based on UMR, CMR and pCMR, and the baseline was the approach termed as "removing missing data" or RMD, in which missing data is simply discarded, easily done when testing for univariate responses. Results for the three sampling times (prelabour, labour and on the neonate) are shown in Tables SM1, 3, and SM2. Comparing RMD outcomes in the three tables, we can conclude that permutation testing is moderately more powerful than traditional approaches, except for some responses where normality assumptions are not met. Our interpretation is that those responses reflect the case of severe deviations from normality illustrated in the power curves of Fig. 2. The tables also illustrate how CMR imputation leads to overoptimistic results in comparison to RMD, situation corrected when applying pCMR. In the remainder of the analyses, we adopt pCMR.

---

[1] Note that the last measurement is not strictly speaking another time point, since the individual change from mother (in the first two time points) to newborn. Yet, our results indicate that this approximation is valid to interpret the data.

**Table 3.** Statistical significance (p-values) in mother's inflammation biomarkers at labour. Bold numbers identify statistically significant values (p-value < 0.05). Responses are described at Table 2. Traditional tests correspond to a t-test when normality assumptions are met or a Wilcoxon test otherwise (for those responses marked with "*"). Missing data approaches are: removing missing data (RMD), unconditional mean replacement (UMR), conditional mean replacement (CMR) and permutational CMR (pCMR).

|  | **Traditional Tests** | | | **Permutation tests** | | | |
| --- | --- | --- | --- | --- | --- | --- | --- |
| **Responses** | **RMD** | **UMR** | **CMR** | **RMD** | **UMR** | **CMR** | **pCMR** |
| **RARRESS** | .44 | .37 | .16 | .29 | .27 | .12 | .23 |
| **LBP** | .21 | .22 | **.02** | **.03** | **.02** | **< .01** | **.03** |
| **IFN-G*** | .41 | .37 | **.04** | **.02** | **.02** | **< .01** | **.01** |
| **IL-10*** | .99 | .92 | .08 | .16 | .13 | **.01** | .07 |
| **IL-23*** | .21 | .22 | **.02** | .21 | .24 | .09 | .21 |
| **IL-6*** | .35 | .25 | **.04** | .73 | .77 | .66 | .78 |
| **IL-8*** | .27 | .23 | **.02** | .11 | .13 | **.03** | .10 |
| **Adip*** | .44 | .34 | .06 | **.04** | **.05** | **.01** | **.03** |
| **Res** | .27 | .23 | **.04** | .08 | .08 | **.02** | .07 |
| **IL-15*** | .90 | .85 | .56 | .65 | .68 | .50 | .68 |
| **VEGF*** | .90 | .25 | .08 | .94 | .93 | .70 | .94 |
| **Lep*** | .27 | .22 | **.02** | .36 | .38 | .21 | .34 |
| **ALR*** | .90 | .92 | .52 | .82 | .81 | .76 | .83 |
| **ARR*** | .90 | .65 | .30 | .78 | .79 | .75 | .83 |

## 6.2 Outlier detection and transformations

The residuals for the 14 auto-scaled responses in the GLM model are visualized with PCA and the D and Q statistics for: the raw data in Fig. 3A, the box-cox transformed data in Fig. 3B, and the rank transformed data in Fig. 3C. The main outliers are both/either situated at the right end or at the top of the figure - indicating they are extreme and/or extraneous to the model, respectively. We observe changes in the distribution and the specific outliers found between non-transformed data in Fig 3A and after the transformations in Figs. 3B and 3C, with higher absolute values of the statistics in the former. Thus, the transformations reduce the relative influence of extreme values on the model calculation, as would be expected from a variance-stabilizing transformation, reducing also the need to discard available data.

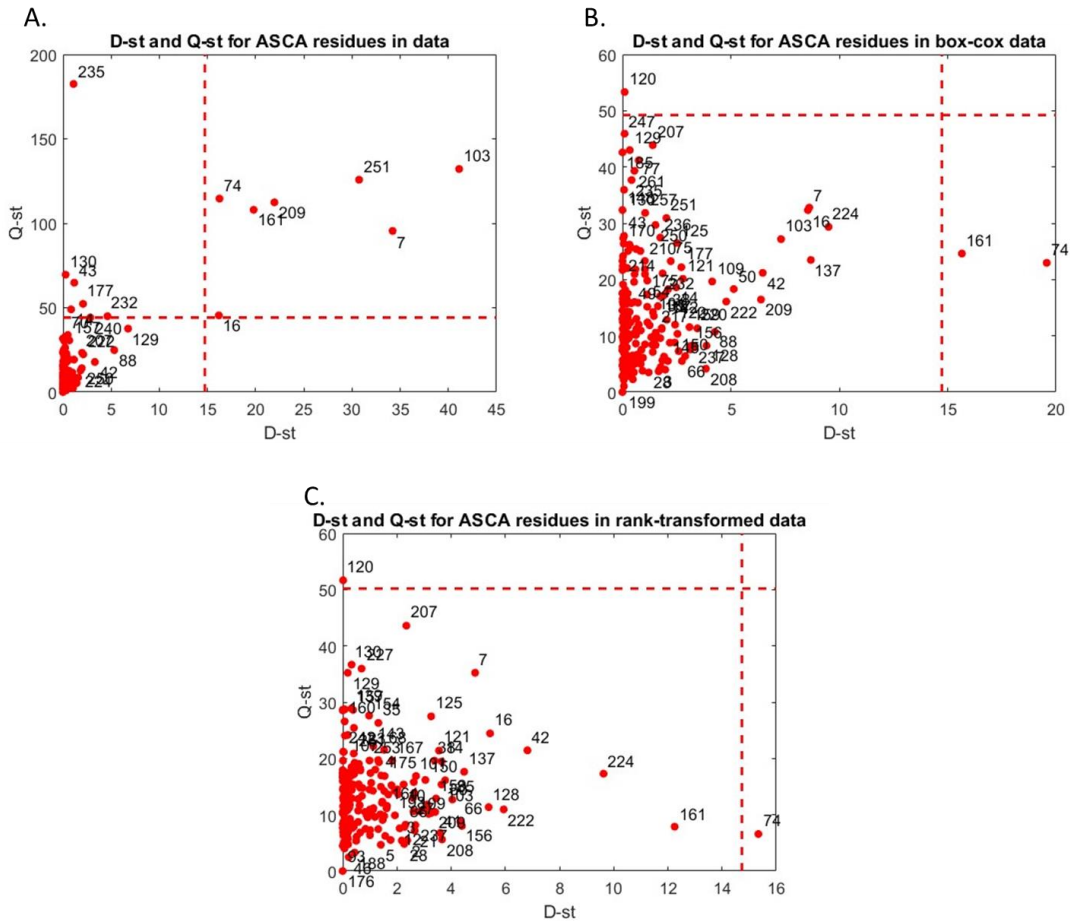

**Fig 3**. D-statistic and Q-statistic of ASCA models' residuals in mother's biomarkers data, as a function of preprocessing.

Inference results for one-way tests for case/control comparisons, including the different transformations and sampling times, are shown in Table 4. We also compare to multi-way ANOVA tests that jointly consider the case/control factor and the time factor, along with their interaction (time factor inference, almost consistently significant, are not presented in the table). Multi-way tests are expected to be a more statistically powerful approach than one-way tests, especially for balanced designs, a situation often not encountered in practice. Given that multi-way ANOVA assumes linearity and heteroscedasticity, and that some responses do not meet such assumptions, we only apply multi-way ANOVA for data transformed with box-cox or the rank transformation. Statistically significant responses are only found in the first time point. We can see that the rank transformation detects the same statistically significant responses as when using either a t-test or a Wilcoxon test in the raw data, the latter in responses where residuals do not meet normality assumptions. Therefore, the rank transformation allows us to simplify the pipeline since t-tests are always used and no normality tests need to be performed. Box-cox transformed data seems to be less statistically powerful. We can also see that multi-way ANOVA tests detect half (2 out of 4) of the significant responses detected in the one-way tests, and they do so for the main case/control factor, rather than in the interaction with time.

Finally in Table 5 we compare multi-way ANOVA tests with multi-way permutation tests based on pCMR, with and without discarding the outliers surpassing to the control limits in Fig. 3. We found coincidental results with and without discarding outliers for the Box-Cox and the rank transformation, but not for the non-transformed data. The rank transformation works also similarly for the permutation testing and ANOVA tests.

**Table 4.** Statistical significance (p-values) for case/control comparisons, including different transformations and sampling times. One-way tests only consider a single time point. Multi-way ANOVA tests consider the case/control and the time factor, along with their interaction. Bold numbers identify statistically significant values (p-value < 0.05). Responses are described at Table 2. Results for non-transformed data (Raw) correspond to a t-test when normality assumptions are met or a Wilcoxon test otherwise (for those responses marked with "*"). Transformed data (either with box-cox (BC) or the rank transform (Rk)) are always based on a t-test.

| Responses | One-way (t-test or Wilcoxon*) | | | | | | | | | Multi-way ANOVA | | | |
|---|---|---|---|---|---|---|---|---|---|---|---|---|---|
| | Prelabour | | | Labour | | | Neonate | | | C/Ctrl Fac. | | Interaction | |
| | Raw | BC | Rk | Raw | BC | Rk | Raw | BC | Rk | BC | Rk | BC | Rk |
| RARR ESS | .26* | .44 | .27 | .44 | .42 | .36 | .84 | .75 | .81 | .65 | .48 | .55 | .56 |
| LBP | **.03** | **.03** | **.03** | .21 | .31 | .21 | .84 | .90 | .81 | **.05** | **< .01** | .55 | .56 |
| IFN-G | .15 | .38 | .50 | .41* | .31 | .36 | .90* | .77 | .93 | .27 | .38 | .90 | .82 |
| IL-10 | .28* | .11 | .27 | .99* | .76 | .97 | .84* | .90 | .96 | .27 | .49 | .55 | .74 |
| IL-23 | **.03*** | .24 | **.03** | .21* | .31 | .21 | .26* | .47 | .69 | **.01** | **< .01** | .55 | .89 |
| IL-6 | .52* | .61 | .63 | .35* | .48 | .38 | .84* | .75 | .81 | .65 | .49 | .55 | .74 |
| IL-8 | .55* | .67 | .62 | .27* | .31 | .35 | .84* | .77 | .81 | .59 | .49 | .56 | .74 |
| Adip | **.03*** | **.03** | **.03** | .44* | .42 | .36 | .50* | .75 | .81 | .38 | .20 | .22 | .28 |
| Res | **.03*** | .06 | **.03** | .27 | .31 | .36 | .84 | .75 | .93 | .27 | .13 | .38 | .56 |
| IL-15 | .41* | .61 | .46 | .90* | .96 | .74 | .90* | .75 | .81 | .71 | .49 | .90 | .92 |
| VEGF | .61* | .67 | .64 | .90* | .96 | .74 | .84* | .79 | .81 | .84 | .74 | .90 | .82 |
| Lep | .55 | .61 | .50 | .27* | .42 | .35 | .49* | .48 | .69 | .84 | .74 | .38 | .28 |
| ALR* | .35* | .44 | .44 | .90* | .96 | .97 | .84* | .75 | .81 | .59 | .48 | .90 | .82 |
| ARR* | .28 | .45 | .44 | .90* | .96 | .74 | .84* | .75 | .81 | .93 | .62 | .68 | .82 |

**Table 5.** Statistical significance (p-values) for case/control comparisons, including different transformations and outlier isolation. Multi-way tests consider the case/control (C) and the time factor, along with their interaction (I). Bold numbers identify statistically significant values (p-value < 0.05). Responses are described at Table 2. Results for non-transformed data (Raw) and transformed data with Box-Cox or the rank transform.

| Responses | Multi-way Permutation tests | | | | | | | | | | | | Multi-way ANOVA | |
|---|---|---|---|---|---|---|---|---|---|---|---|---|---|---|
| | Raw | | Raw no outliers | | Box-Cox | | Box-Cox no outliers | | Rank | | Rank no outliers | | Rank | |
| | C | I | C | I | C | I | C | I | C | I | C | I | C | I |
| RARRESS | .90 | .31 | .95 | .44 | .64 | .63 | .64 | .63 | .59 | .49 | .59 | .49 | .48 | .56 |
| LBP | **.02** | .57 | .05 | .40 | **.04** | .67 | **.03** | .52 | **< .01** | .55 | **< .01** | .55 | **< .01** | .56 |
| IFN-G | **.04** | .26 | **.03** | .33 | .28 | .90 | .31 | .95 | .35 | .82 | .35 | .82 | .38 | .82 |
| IL-10 | .06 | .30 | .98 | .53 | .26 | .57 | .26 | .51 | .51 | .78 | .51 | .78 | .49 | .74 |
| IL-23 | .86 | 1.0 | .41 | .92 | **< .01** | .65 | **< .01** | .53 | **< .01** | .87 | **< .01** | .87 | **< .01** | .89 |
| IL-6 | .91 | .95 | .32 | .61 | .69 | .59 | .69 | .56 | .47 | .87 | .47 | .87 | .49 | .74 |
| IL-8 | .95 | .44 | .98 | .93 | .57 | .58 | .44 | .50 | .46 | .82 | .46 | .82 | .49 | .74 |
| Adip | .91 | .08 | .92 | .35 | .36 | .17 | .25 | .12 | .17 | .34 | .17 | .34 | .20 | .28 |
| Res | .92 | .37 | .38 | .68 | .28 | .32 | .25 | .29 | .11 | .57 | .11 | .57 | .13 | .56 |
| IL-15 | .91 | .95 | 1.0 | .87 | .70 | .95 | .48 | .89 | .51 | .93 | .51 | .93 | .49 | .92 |
| VEGF | .93 | .95 | 1.0 | .63 | .83 | .93 | .88 | .85 | .72 | .81 | .72 | .81 | .74 | .82 |
| Lep | .92 | .31 | .99 | .47 | .84 | .28 | .83 | .31 | .75 | .19 | .75 | .19 | .74 | .28 |
| ALR* | 1.0 | 1.0 | .30 | .75 | .58 | .89 | .41 | .95 | .52 | .82 | .52 | .82 | .48 | .82 |
| ARR* | .98 | .46 | 1.0 | .40 | .93 | .66 | .91 | .62 | .61 | .81 | .61 | .81 | .62 | .82 |

# 7 Discussion and suggestions of good practices

Results presented in both simulations and real data demonstrate the complexity of the problem treated. Yet, we conclude that real data agrees with the trends observed in the simulations, which allows us to provide the following rules of thumb that are also consistent with common wisdom in ANOVA and related analysis tools.

Missing data imputation for univariate responses can be done by discarding the missing elements. When multiple responses are treated jointly, like in ASCA, this may lead to discarding incomplete observations, which is a suboptimal use of data. An alternative approximation with good inferential behaviour (similar in our results to discarding missing elements) is to use a missing data imputation scheme in permutation testing, both outside and inside the permutation loop, like pCMR.

As it is well-known, assumptions about normality and heteroscedasticity in residuals are relevant in t-test and ANOVA tests. We illustrated several situations in which the failure to meet such assumptions affected the inference results. Permutation tests were found to be more robust to deviations from the assumptions, but not immune to them. In particular, we found the combination of permutation testing and the rank transformation very effective. The rank transformation is also, as expected, very robust to outliers. Yet, the rank transformation is very aggressive, and visualizations of the data and residuals in ANOVA and ASCA may change to a large extent. Our suggestion is to repeat the ANOVA/ASCA pipeline i) with the raw data (or a milder transformation like Box-Cox) and with outliers detection and isolation, ii) with the rank transformed data. If both pipelines are consistent in terms of statistical inference, the analyst may prefer i) for a better interpretation of post-hoc tests/PCA visualizations. Otherwise, working on the rank-transformed data (either for all responses or for selected ones) may be preferable.

# 8 Conclusions

Permutation testing is an alternative to traditional parametric tests in Analysis of Variance (ANOVA) which allows for more flexibility in the distribution of the residuals. It is also the facto standard for the statistical inference in ANOVA Simultaneous Component Analysis (ASCA) models, arguably one of the most popular multivariate extensions of ANOVA. The effect of missing values has a profound impact on what experimental factors are indicated as significant in ANOVA and ASCA models. Mean-replacement methods for missing value imputation can be performed either unconditionally (i.e. agnostically to the known experimental factors) or conditionally (i.e. as a function of the known experimental factors). These methods are prone to increasing the risks of types I & II errors, respectively. An additional consideration depends on whether the model residuals are distributed normally and with constant variance. In cases where this is not true, the logarithmic (i.e. Box-Cox) or rank transformations can be used.

In this paper, we study the effect of missing data imputation, outliers detection and isolation, and transformations in the analysis of parametric ANOVA, permutation-based ANOVA and ASCA in both simulated and real data. We provide reasonable rules to perform these steps in real problems. A "conditional mean replacement by cell" method via permutation testing is proposed to handle missing data. An interesting combination of permutation testing with the rank transformation is also highlighted to validate inference results in the presence of outliers and complex distributions.

**Acknowledgements**

This work was partly supported by the Agencia Andaluza del Conocimiento, Regional Government of Andalucıa, in Spain and ERDF (European Regional Development Fund) funds through project B-TIC-136-UGR20; by the Agencia Estatal de Investigación in Spain, MCIN/AEI/ 10.13039/501100011033, grant No PID2020-113462RB-I00; and by the Institute of Health Carlos III (ISCIII) PI17/01215. Michael Sorochan Armstrong was funded through the Marie Skłodowska-Curie Actions Program, project: MAHOD-101106986.

# Supplementary material

**Table SM1.** Statistical significance (p-values) in mother's inflammation biomarkers at pre-labour. Bold numbers identify statistically significant values (p-value < 0.05). Responses are described at Table 2. Traditional tests correspond to a t-test when normality assumptions are met or a Wilcoxon test otherwise (for those responses marked with "*"). Missing data approaches are: removing missing data (RMD), unconditional mean replacement (UMR), conditional mean replacement (CMR) and permutational CMR (pCMR).

|  | **Traditional Tests** | | | **Permutation tests** | | | |
|---|---|---|---|---|---|---|---|
| **Responses** | **RMD** | **UMR** | **CMR** | **RMD** | **UMR** | **CMR** | **pCMR** |
| **RARRESS*** | .26 | .21 | .06 | .32 | .33 | .21 | .30 |
| **LBP** | **.03** | **.04** | **.01** | **.01** | **.01** | **< .01** | **.01** |
| **IFN-G** | .15 | .16 | .06 | .05 | .05 | **.04** | .05 |
| **IL-10*** | .28 | .30 | .15 | **.01** | **.01** | **.01** | **.02** |
| **IL-23*** | **.03** | **.04** | **.01** | .80 | .83 | .71 | .80 |
| **IL-6*** | .52 | .37 | .06 | .91 | .82 | .77 | .88 |
| **IL-8*** | .55 | .37 | **.01** | .82 | .90 | .97 | .78 |
| **Adip*** | **.03** | .06 | **.01** | **< .01** | **< .01** | **< .01** | **< .01** |
| **Res*** | **.03** | **.04** | **.01** | **< .01** | **.01** | **< .01** | **< .01** |
| **IL-15*** | .41 | .43 | .29 | .86 | .89 | .89 | .83 |
| **VEGF*** | .61 | .64 | **.02** | .68 | .90 | .87 | .77 |
| **Lep** | .55 | .55 | .47 | .51 | .51 | .47 | .53 |
| **ALR*** | .29 | .30 | .15 | .4 | .41 | .34 | .37 |
| **ARR*** | .28 | .30 | .15 | .16 | .14 | .12 | .15 |

**Table SM2.** Statistical significance (p-values) in mother's inflammation biomarkers in newborns. Bold numbers identify statistically significant values (p-value < 0.05). Responses are described at Table 2. Traditional tests correspond to a t-test when normality assumptions are met or a Wilcoxon test otherwise (for those responses marked with "*"). Missing data approaches are: removing missing data (RMD), unconditional mean replacement (UMR), conditional mean replacement (CMR) and permutational CMR (pCMR).

|  | **Traditional Tests** | | | **Permutation tests** | | | |
|---|---|---|---|---|---|---|---|
| **Responses** | **RMD** | **UMR** | **CMR** | **RMD** | **UMR** | **CMR** | **pCMR** |
| **RARRESS** | .84 | .90 | .24 | .35 | .34 | .18 | .30 |
| **LBP** | .84 | .90 | .26 | .62 | .63 | .47 | .63 |
| **IFN-G**\* | .90 | .98 | .85 | .91 | .92 | .87 | .90 |
| **IL-10**\* | .84 | .93 | .72 | .74 | .74 | .62 | .74 |
| **IL-23**\* | .26 | .56 | **< .01** | .24 | .36 | .05 | .21 |
| **IL-6**\* | .84 | .99 | **.07** | .30 | .27 | .35 | .31 |
| **IL-8**\* | .84 | .90 | .21 | .32 | .26 | .21 | .33 |
| **Adip**\* | .50 | .80 | .07 | .11 | .10 | **.02** | .09 |
| **Res** | .84 | .90 | .28 | .33 | .36 | .29 | .36 |
| **IL-15**\* | .90 | .98 | .84 | .84 | .87 | .78 | .84 |
| **VEGF**\* | .84 | .90 | .06 | .35 | .63 | **.01** | .39 |
| **Lep**\* | .49 | .80 | **.01** | .18 | .16 | .06 | .16 |
| **ALR**\* | .84 | .90 | .17 | .64 | .76 | .57 | .66 |
| **ARR**\* | .84 | .90 | .23 | .18 | .19 | .12 | .18 |